\documentclass{article}
\usepackage{spconf}
\usepackage{ifpdf}
\usepackage{cite}
\usepackage{graphicx}
\usepackage{amsmath}
\usepackage{algorithmic}
\usepackage{array}
\usepackage{blindtext}
\usepackage{epstopdf}
\usepackage{bm}
\usepackage{enumerate}
\usepackage{amsbsy}
\usepackage{amssymb}
\usepackage{amsthm}
\usepackage{amscd}
\usepackage{subfigure}
\usepackage{color}
\usepackage{booktabs}
\usepackage{multirow}
\usepackage{hhline}
\usepackage{makecell}
\usepackage{bbold}
\usepackage{url}
\usepackage{tabularx}
\usepackage{cellspace}
\usepackage{boldline}
\usepackage[]{algorithm2e}
\usepackage[inline]{enumitem}

\def\bomg{{\boldsymbol{\omega}}}

\def\c1{{\textcircled{a}}}

\def\bp{{\boldsymbol{p}}}
\def\bq{{\boldsymbol{q}}}

\def\pP{\mathbb{P}}
\def\pR{\mathbb{R}}

\title{Comprehensive Personalized Ranking Using One-Bit Comparison Data}
\name{Aria~Ameri*, Arindam~Bose, Mojtaba~Soltanalian\thanks{* Corresponding author (e-mail: \textit{aameri2@uic.edu}). This work was supported in part by U.S. National Science Foundation Grant CCF-1704401.}
\address{Department of Electrical and Computer Engineering\\University of Illinois at Chicago\\Chicago, USA}}
\begin{document}

\maketitle

\begin{abstract}
The task of a personalization system is to recommend items or a set of items according to the users' taste, and thus predicting their future needs. In this paper, we address such personalized recommendation problems for which one-bit comparison data of user preferences for different items as well as the different user inclinations toward an item are available. We devise a comprehensive personalized ranking (CPR) system by employing a Bayesian treatment. We also provide a connection to the learning method with respect to the CPR optimization criterion to learn the underlying low-rank structure of the rating matrix based on the well-established matrix factorization method. Numerical results are provided to verify the performance of our algorithm.
\end{abstract}
\vspace{-1.5em}
\section{Introduction}
\label{sec:intro}
Content recommendation is one of the most vital tasks emerging in various online systems \cite{Das:2007:GNP:1242572.1242610, 1167344, Sharma:2013:PLR:2507157.2507175, adomavicius2005toward} and mobile applications \cite{doi:10.1142/S0219622013500077, PAN2015173}. Since the advent of the Netflix challenge \cite{Netflix}, study of recommendation systems, also known as personalization systems, has been an active research area. The ultimate goal of a recommendation system is to endorse items based on different parameters, such as the preference of a user, the relevance of an item and the quality of its content, contemporary trend, among others  \cite{Rendle:2009:BBP:1795114.1795167}. Recommendations can be in the form of predicting the user preferences for an unobserved item or a list of top relevant items. In order to form such recommendations, personalization systems generally rely on explicit preferences of users based on readily available evidence such as ratings, votes or reviews. However, in practice, implicit feedback, such as clicks on a link or views of a video may also contribute to define user preferences. In many cases, the implicit feedback can only be observed in a rather obscure binary manner, e.g. a `one' can signify that the user has an interest in an item, while a `zero' may represent a disliked or merely an unobserved item. Note that user ratings can be broadly subjective and are prone to change over time, and thus, may not reflect an absolute predisposition on the item. Moreover, in a strictly quantized rating system such as a 5-star based system, the choices to show due appreciation for the items are very limited and many products can be identified by an identical rating while their ranking order is somewhat different in the mind of the user.

These central issues associated with recommendation systems can be partly mitigated by using one-on-one comparisons for item pairs. In this way, users can be more expressive in choosing between two items without putting much effort, rather than just providing independent opinions about one item without the presence of any arbitrary frame of reference. In such contexts, implicit feedback can also be utilized to recommend better items. On the other hand, user-user dependency for an item, which may be inferred from the same user's voting pattern, can also provide more information to the recommendation system. For example, it can indicate a stronger inclination of one group of users toward an item than another group, which is very common in practical scenarios. 

In the literature, the problem of personalization has been approached from many different angles based on the context of the recommendation process. One conventional method utilized for recommendation systems is k-nearest neighbors (kNN) in order to perform item-item or user-user collaborative filtering employing the correlation between item or user pairs, respectively \cite{Deshpande:2004:ITN:963770.963776}. Popular methods like Matrix Factorization (MF) has also been considered in several studies \cite{1565702, bose2018low}. The problem of estimating the rating matrix via the MF method is especially relevant in the low-rank setting where the matrix can be represented as the product of two low-rank matrices that contain the latent features of the corresponding users and items, respectively. In a collaborative filtering context, \cite{bose2018low} presents a novel approach to solve the non-convex problem of recovering the underlying low-rank structure of the rating matrix based on one-bit comparisons between pair of items using an alternating optimization algorithm. A novel method of personalized ranking system has been proposed in \cite{Rendle:2009:BBP:1795114.1795167}, in which the authors present a probabilistic argument based on a Bayesian analysis using item-item comparison data as a prior. 

In this paper, we present a Bayesian framework to address the recommendation problem through a Comprehensive Personalized Ranking (CPR) system, which not only utilizes the one-bit item-item preference of a user, but also exploits the implicit inclination of different users towards an item. We provide a stochastic-gradient based approach to learn the system parameters. We also discuss how the aforementioned algorithm can be utilized to estimate the user/item latent features using MF. \nocite{ShahinAsilomar, NaveedAsilomar} In fact, MF will be used to jump-start CPR.

The remainder of the paper is organized as follows. Section \ref{sec:prob} presents the problem formulations along with a Bayesian treatment of the data. We also provide an MF based approach to estimate the low-rank rating matrix. Numerical results are presented in Section \ref{sec:num}. Finally, Section \ref{sec:con} concludes the paper.
\vspace{5pt}

\textit{Notation:} We use bold lowercase letters for vectors and uppercase letters for matrices. $(\cdot)^T$ denotes the vector/matrix transpose.

\vspace{-1.5em}
\section{Problem Formulation}
\label{sec:prob}

Let $U$ denote the set of all users and $I$ denote the
set of all items. We let $\Omega$ represent the internal system parameter which could be viewed as a user/item latent feature matrix. The users are asked to compare two items based on their best judgment at the time. From such set of queries, we can clearly infer the users' inclination toward a specific item. Such data can also be implicitly collected from user activity associated with an item, e.g. records of clicks on a link, or interest for a video content. The goal of the recommendation system is to provide effective personalized recommendations based on observed user-user and item-item comparison data, denoted by $>_m \subset U^2$ and $>_u \subset I^2$, respectively, where $>_m$ and $>_u$ denote the observed user-user comparisons for item $m$ and the item-item comparisons by user $u$, respectively. Note that the operator $>_m$ has to meet the following three properties \cite{Rendle:2009:BBP:1795114.1795167}:
\begin{align*}
	totality&: i\neq_m j \implies i >_m j \vee j >_m i:\forall~i,j \in U\\
	antisymmetry&: i >_m j \wedge j >_m i \implies i=_m j:\forall~i,j \in U\\
	transitivity&: i >_m j \wedge j >_m k \implies i >_m k:\\
	&\qquad\qquad\qquad\qquad\qquad\qquad\forall~i,j,k \in U.
\end{align*}
Particularly, the transitivity property mentioned above is used to extract additional information from the comparison database. In a similar manner, $>_u$ has to meet the same three properties as well.

In this paper, as mentioned earlier, instead of using the exact rating values, we use the user-user and item-item one-bit comparison data in order to design an efficient personalized ranking/recommendation system.
In the following section, we present a Bayesian treatment of the data to solve the problem of Comprehensive Personalized Ranking (CPR) by using the likelihood function for $\pP\{i>_m j, k >_u l~|\Omega\}$ and the prior probability of the system parameters $\pP\{\Omega\}$.

\subsection{Analysis of the Likelihood and Posterior Functions}
\label{subsec:likelihood}

Note that we are interested in the maximization of the following posterior probability $\pP\{\Omega | >_m, >_u\}$ where $\Omega$ denotes the matrix of system parameters for all $i,j \in U$ and $k,l \in I$:
\begin{align}\label{eq:init}
	\pP\{\Omega| >_m, >_u\} = \alpha \cdot \pP\{>_m, >_u|\Omega\}~\pP\{\Omega\},
\end{align} 
where $\alpha$ is a constant term independent of $\Omega$. Here we assume that, when $\Omega$ is given (equivalent to us having the rating matrix), not only the ordering of each pair of items becomes independent of rest of the orderings, but also no two users can any longer influence the votes of each other. Hence, the likelihood function, $\pP\{i>_m j, k >_u l~|\Omega\}$ can not only be represented as a multiplication of two independent probability distributions, i.e. $ \label{eq:likelihood}
	\pP\{>_m, >_u|\Omega\} = \pP\{>_m |\Omega\}\pP\{ >_u|\Omega\} $,
but also following the totality and antisymmetry property,
\begin{align}
	\pP\{>_m |\Omega\} &= \prod_{(i,j) \in D_m}{\pP\{i >_m j|\Omega\}}, \nonumber\\
	\pP\{>_u |\Omega\} &= \prod_{(k,l) \in D_u}{\pP\{k >_u l|\Omega\}},
\end{align}
hold true, for each independent item/user, $m \in I$ and $u \in U$, where $D_m$ and $D_u$ are the set of users who rated item $m$, and the set of observed or rated items for user $u$, respectively.

Now, in order to form personalization based on users' own choices, we define the individual probability functions for the case where user $u$ likes item $k$ over item $l$, and in a similar fashion, user $i$ has a stronger inclination toward item $m$ than user $j$, based on the system parameters:
\begin{align}
	\pP\{i >_m j|\Omega\} &\triangleq f(c_m, \hat{x}_{ijm}(\Omega)), \nonumber\\
	\pP\{k >_u l|\Omega\} &\triangleq f(c_u, \hat{x}_{ukl}(\Omega)),
	\label{eq:likelihoods}
\end{align}
where we define the estimators $\hat{x}_{ijm}(\Omega)$ and $\hat{x}_{ukl}(\Omega)$ as the difference between two individual estimators of the actual ratings:
\begin{align}\label{eq:mfd}
	\hat{x}_{ijm}(\Omega) &\triangleq \hat{x}_{im}(\Omega) - \hat{x}_{jm}(\Omega), \nonumber \\
	\hat{x}_{ukl}(\Omega) &\triangleq \hat{x}_{uk}(\Omega) - \hat{x}_{ul}(\Omega),
\end{align}
and where $\hat{x}_{pq}(\Omega)$, for $p \in U$ and $q \in I$, represents the estimate of the actual rating of user $p$ for item $q$. According to item response theory \cite{embretson2013item}, the function $f(\cdot)$ is a mapping from the (comparative) rating space to a probability, defined as $ f(c,x) \triangleq \frac{1}{2} + \frac{1}{2}\tanh(cx). $

\textit{Remark: }The constants $c_m$ and $c_u$ in \eqref{eq:likelihoods} are the model specific parameters which determine the relevance of the query about item $m$ and user $u$, respectively, and their values can be found empirically. The bigger the constant $c$, the steeper the function $f(c,x)$ becomes around the origin, and thus, the more reliability we put on the estimated ratings. \hfill$ \blacksquare $

In the following, we discuss how these user/item entity specific functions can be estimated for a given $\Omega$ using well established methods such as matrix factorization (MF).

\vspace{-1em}
\subsubsection{Derivation of the user/item entity specific functions $\hat{x}_{ijm}(\Omega)$ and $\hat{x}_{ukl}(\Omega)$ using MF}
\label{subsubsec:mf}

Note that the problem of estimating $\hat{x}_{im}(\Omega)$ in \eqref{eq:mfd}, for all $i\in U$ and all $m\in I$, is equivalent to estimating the rating matrix $X\in\pR^{|U|\times |I|}$. This task has been successfully handled before by using the MF method, where the matrix $X$ can be approximated by the product of two low-dimensional matrices $P\in\pR^{|U| \times r}$ and $Q\in\pR^{|I| \times r}$ as $\hat{X} = PQ^T$, and where $r$ is the number of features that characterizes items and users, or in other words, the rank of the approximation \cite{bose2018low}. Note that the corresponding rows of $P$ and $Q$, denoted by $\{\bp_u^T\}_{u=1}^{|U|}$ and $\{\bq_m^T\}_{m=1}^{|I|}$, can be used to constitute a modified system parameter matrix $\Omega = [P^T|Q^T]$, where the rating estimates are:
\vspace{-1.5em}\begin{align*}
	\hat{x}_{im} &\triangleq \langle \bp_i,\bq_m\rangle = \bp_i^T\bq_m = \sum_{t=1}^{r}{p_{it}q_{tm}}, \nonumber \\
	\hat{x}_{uk} &\triangleq \langle \bp_u,\bq_k\rangle = \bp_u^T\bq_k = \sum_{t=1}^{r}{p_{ut}q_{tq}}.
\end{align*}
It is now straightforward to verify that $\hat{x}_{ijm} = (\bp_i-\bp_j)^T\bq_m$ and $\hat{x}_{ukl} = \bp_u^T(\bq_k-\bq_l)$. Another method to estimate these entity specific functions using a non-convex optimization method based on MF can be found in \cite{bose2018low}.

\vspace{-1em}
\subsection{Analysis of the Prior Function}
\label{subsec:prior}
\vspace{-.4em}
In this section, we introduce the general prior density function $\pP\{\Omega\}$ of the system parameters $\Omega= [P^T|Q^T]=[\bomg_1 \cdots \bomg_N]$ as independent normalized zero-mean multivariate normal random variables with known covariance matrices $\{\Sigma_n\}_{n=1}^{N}$ where $N$ is the number of parameter vectors in $\Omega$. For $\Omega$, as described in subsection \ref{subsubsec:mf}, $N$ would be equal to $|U|+|I|$. Given these assumptions, $\pP\{\Omega\}$ can be formulated as,
\begin{align}
	\pP\{\Omega\} = \frac{1}{(2\pi)^{\frac{N}{2}}\prod_n{|\Sigma_n|^{\frac{1}{2}}}}\exp\left\lbrace-\frac{1}{2}\sum_n{\bomg_n^T\Sigma_n^{-1}\bomg_n}\right\rbrace.
\end{align}

\vspace{-2em}
\subsection{Comprehensive Personalized Ranking (CPR)}
\label{subsec:cpr}
With the above discussion in mind, we can rewrite our likelihood function in (\ref{eq:likelihood}) as,
\begin{align}
	&{\pP\{>_m, >_u|\Omega\}} = \nonumber\\
	&\prod_{m = 1}^{|I|}\prod_{(i,j) \in D_m}{f(c_m, \hat{x}_{ijm}(\Omega))} \times \nonumber
	\prod_{u = 1}^{|U|}\prod_{(k,l) \in D_u}{f(c_u, \hat{x}_{ukl}(\Omega))}.
\end{align}
Furthermore, we finally formulate our personalization problem in terms of maximizing the below criterion with respect to (w.r.t.) $\Omega$:
\vspace{-.6em}
\begin{eqnarray}\label{eq:cpr}
	\text{CPR}~~~\triangleq& &\ln{\pP\{\Omega | >_m, >_u\}} \nonumber \\
	\simeq& &\ln{\pP\{>_m, >_u|\Omega\}\pP\{\Omega\}} \nonumber \\
	\simeq& &\sum_{m}\sum_{(i,j) \in D_m}{\ln f(c_m, \hat{x}_{ijm}(\Omega))} \nonumber \\
	&+ &\sum_{u}\sum_{(k,l) \in D_u}{\ln f(c_u, \hat{x}_{ukl}(\Omega))} \nonumber \\
	&-&\frac{1}{2}\sum_n{\bomg_n^T\Sigma_n^{-1}\bomg_n}
\end{eqnarray}

\vspace{-2em}
\subsection{Learning the CPR}
\label{sub:learnCPR}
We begin by calculating the gradient of CPR w.r.t. the model parameter $\Omega$, denoted as $\frac{\partial}{\partial\Omega}\text{CPR}$, through considering each term at the right hand-side of (\ref{eq:cpr}) individually. It can be easily verified that, 
\vspace{-.5em}
\begin{align}
	\frac{\partial}{\partial\Omega}{\ln f(c, \hat{x})}=c(1-\tanh(c \hat{x}))~\frac{\partial}{\partial\Omega}\hat{x}.
\end{align}
As one can observe from (\ref{eq:cpr}), we also need to know the gradients of $\hat{x}_{ijm}(\Omega)$ and $\hat{x}_{ukl}(\Omega)$, w.r.t. each system parameter $\bomg=\{\omega_t\}_{t=1}^r$, and can be given for (\ref{eq:mfd}) as,
\vspace{-.5em} 
\begin{align}
	\frac{\partial}{\partial\Omega}\hat{x}_{ijm}=\left\lbrace
	\begin{array}{cl}
	(p_{it}-p_{jt}), & \omega_t=q_{tm},\\
	q_{tm}, &\omega_t=p_{it},\\
	-q_{tm}, &\omega_t=p_{jt},\\
	0, &\text{otherwise},
	\end{array}\right.
\end{align} 
and $\frac{\partial}{\partial\Omega}\hat{x}_{ukl}$ can be derived in a similar fashion. The gradient of the last term in (\ref{eq:cpr}) can be derived as,
\begin{align}
	\frac{\partial}{\partial\Omega}\frac{1}{2}\sum_n{\bomg_n^T\Sigma_n^{-1}\bomg_n} = [\Sigma_1^{-1}\bomg_1 \cdots \Sigma_N^{-1}\bomg_N].
\end{align}
Note that in a case where $\{\Sigma_n\}_{n=1}^{N} = \Sigma$, the latter expression can be simplified as $\Sigma^{-1}\Omega$.

In light of the above, the system parameter $\Omega$ can be updated by employing a (stochastic) gradient descent method:
\begin{align}
	\Omega_{new} \leftarrow \Omega - \mu \frac{\partial}{\partial\Omega}\text{CPR},
\end{align}
where $\mu$ represents a proper positive learning rate. 

\vspace{-0.5em}
\section{Numerical Results}
\label{sec:num}
\vspace{-1em}

\begin{figure*}[t]
	\centering
	\subfigure[]{
		\includegraphics[width=0.38\linewidth]{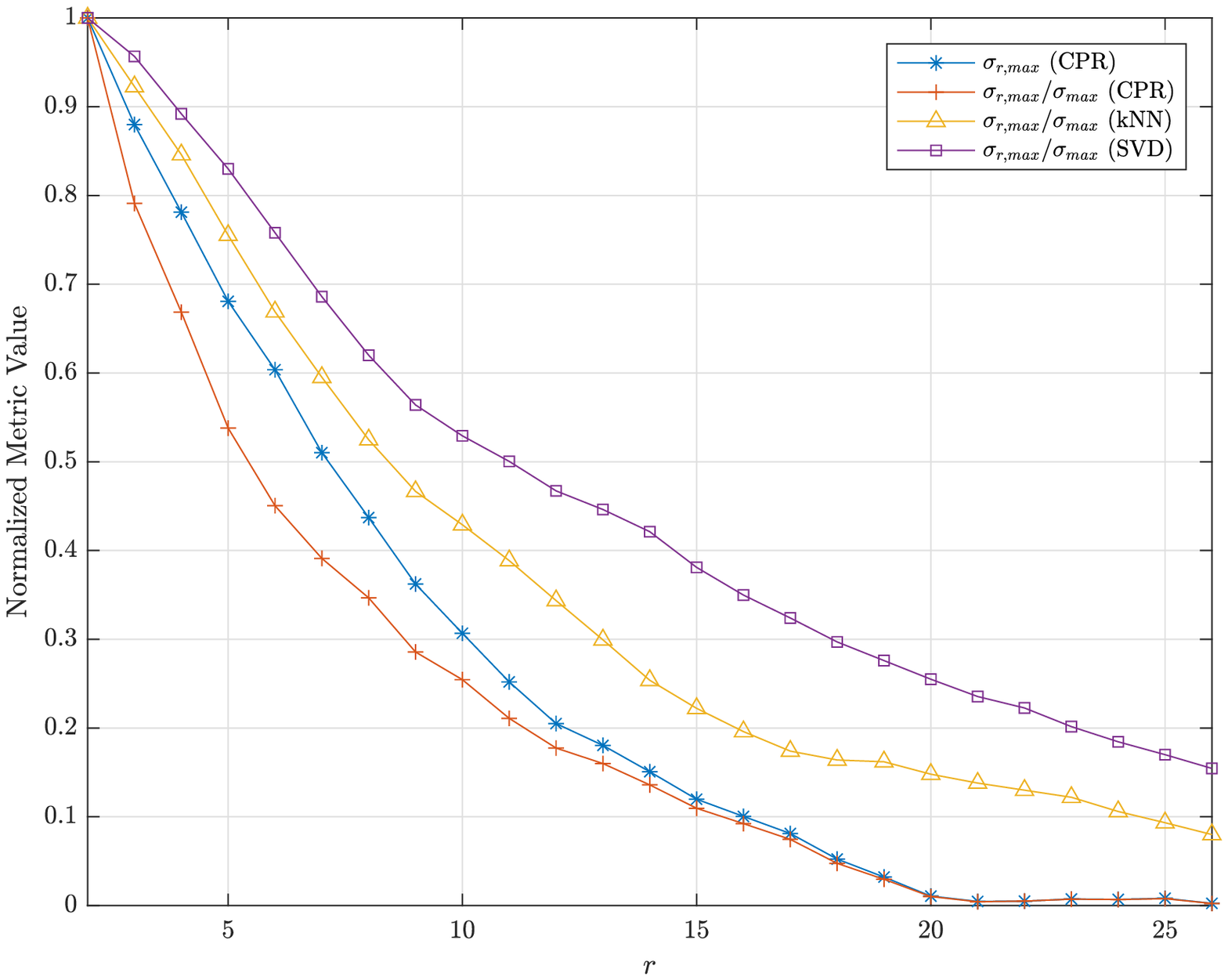}
	}
	\subfigure[]{
		\includegraphics[width=0.38\linewidth]{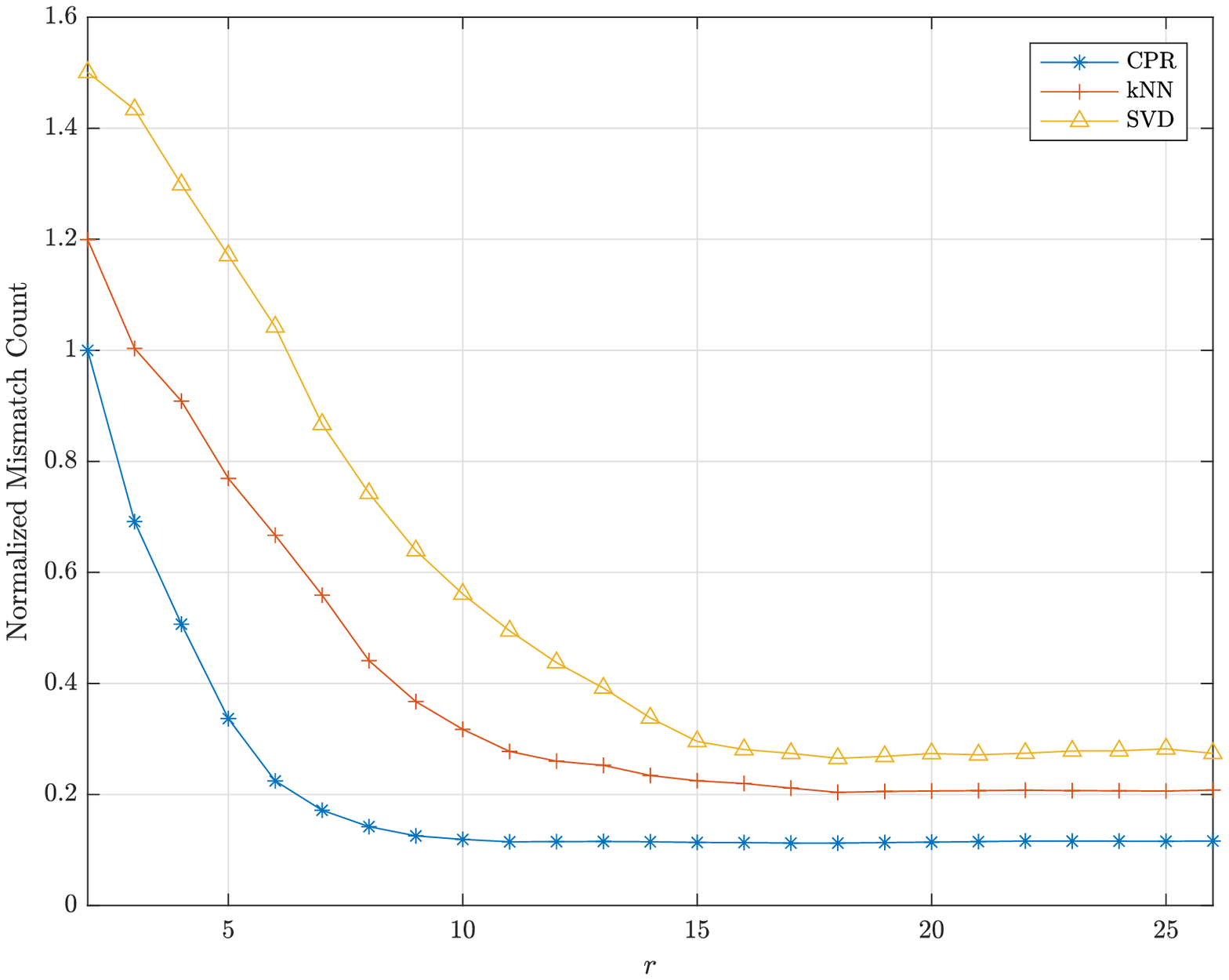}
	}
	\vspace{-1em}
	\caption{The results for different algorithms: (a) normalized values of various metrics on the recovered rating matrices versus the expected rank $r$, (b) the normalized number of mismatches between the original comparison data and the comparisons made from the recovered data for $CPR$, $kNN$ and $SVD$.}
	\label{fig:metrics}
	\vspace{-1em}
\end{figure*}

In this section, we show that the proposed method can efficiently recover the original low-rank rating matrix in terms of the comparison relationships through learning the underlying low-rank structure. As a byproduct, it offers the missing information in the rating matrix, which is the desired objective of a good recommendation system: recommending new items to users based on their own interests and the interests of other like-minded users. \cite{deshpande2004item}.

For simulations, we use a portion of the publicly available MovieLens datasets containing movie ratings---about 600 ratings given by 40 users judging 60 movies on a scale between 1 and 5 \cite{Harper:2015:MDH:2866565.2827872}. We start by converting the rating matrix to comparison data and these data are stored in a memory-efficient way. In order to train the system based on a large amount of observed data, 
 we resort to the stochastic gradient descent method and mini-batch learning \cite{hinton2012neural}, which is proven to be robust enough to provide near optimal results asymptotically \cite{shamir2013stochastic}. Accordingly, we utilize the stochastic gradient method with a batch size of $32$ for simulation purposes.

According to the analysis in Section~\ref{sec:prob}, it is clear that the proposed method relies on processing the data in an $ r $-dimensional space. Additionally, as many users tend to show shared interest in only specific subsets of items, the rating matrix is, naturally, low-rank. In order to concur with this behavior of the rating matrix and to use the many advantages of low-rank matrices such as decrease in the computation and storage burdens, we resort to finding the best low-rank matrix approximation for the given data. A natural metric for determining the rank of the original rating matrix, denoted by $r_X$, can be to look at its $r$ largest singular values of the recovered matrix. When the expected rank $r$ is smaller than $r_X$, the algorithm cannot allocate all the information in an $r$-dimensional space. As a result, the updated $(r+1)$-th dimension will have a considerable amount of information that cannot be ignored. Alternatively, when $r$ is larger than $r_X$, the algorithm transcribes most of the recovered information in an $r_X$-dimensional space and places little to no information in the remaining $r-r_X$ dimensions. Consequently, one can use the ratio of the $r$-th largest to the largest singular value of the recovered matrix as a metric to determine the true rank. According to the above analysis, the mentioned ratio should drop drastically as soon as $r$ gets greater than $r_X$, enabling us to determine $r_X$.

In Fig.~\ref{fig:metrics}~(a), the average performance of the proposed method, denoted by $CPR$, is quantified w.r.t.  \begin{enumerate*}[label=(\roman*)]
	\item the $r$-th largest singular value of the recovered rating matrix, denoted by $\sigma_{r,max}$, and
	\item the ratio of $r$-th largest to the largest singular value of the recovered rating matrix, denoted by $\sigma_{r,max}/\sigma_{max}$,
\end{enumerate*} where all traces are normalized by their first elements. It can be seen that the knee point occurs at $r=20$, which means that the best rank for the recovery is $20$. Similar phenomenon can be noticed for $\sigma_{r,max}/\sigma_{max}$, as the added dimensions do not bear as much information as before, verifying the correctness our intuition. It can also be seen that the performance of $CPR$, in terms of finding the best low-rank matrix by allocating more information in smaller dimensions, is better than the other state of the art methods such as $kNN$ and $SVD$ \cite{Deshpande:2004:ITN:963770.963776, adomavicius2005toward}. Fig.~\ref{fig:metrics}~(a) depicts that $CPR$ stores the information in a rank-$ 20 $ matrix while other methods still have information to be stored in the matrix of rank above $ 20 $.

Furthermore, in Fig.~\ref{fig:metrics}~(b) we compare the number of mismatches between the original comparison data and the comparisons made from the recovered data for $CPR$, $kNN$, and $SVD$, all normalized to the value of mismatches in rank-$ 2 $ recovered matrix by $CPR$. It should be noted that the number of mismatches for the proposed method is less than that of other methods and remains virtually the same for $r \geq 10$, This further verifies superior performance of $CPR$ as compared to other methods. Thus, it achieves the minimum number of mismatches among others while finding the best low-rank matrix approximation.



\vspace{-1em}
\section{Conclusion and Future Work}
\label{sec:con}
\vspace{-1em}
In this paper, we studied a new optimization framework based on one-bit preference comparison data to develop a Comprehensive Personalized Ranking (CPR) system. The algorithm relies on a Bayesian treatment of the data, and maximizes the posterior probability of the system parameters. A learning model w.r.t. the aforementioned optimization problem is provided. Several numerical results were provided to show the effectiveness of the algorithm. The study of the impact of the rating matrix size on the projected rank would be an interesting future research avenue as the projected rank of a matrix significantly controls the storage and computational efficiency of the algorithm. 

\bibliographystyle{IEEEbib}
\bibliography{refs}

\end{document}